\documentclass{svproc}
\usepackage{graphicx}%
\usepackage{multirow}%
\usepackage{amsmath,amssymb,amsfonts}%
\usepackage{mathrsfs}%
\usepackage[title]{appendix}%
\usepackage{xcolor}%
\usepackage{textcomp}%
\usepackage{manyfoot}%
\usepackage{booktabs}%
\usepackage{algorithm}%
\usepackage{algorithmicx}%
\usepackage{algpseudocode}%
\usepackage{listings}%
\usepackage{svg}%
\usepackage{url}

\usepackage[colorinlistoftodos,prependcaption]{todonotes}

\begin{document}
\mainmatter              
\title{Unravelling Nature's Models for Transportation Network: Considering a Biomimicry Framework}
\titlerunning{Unravelling Nature's Models for Transportation Network}  
%
\author{Sofiane Madmar\inst{1,2} \and Didier Josselin\inst{1} \and
Olivier Blight\inst{2} \and Vincent Labatut\inst{3} \and Christophe Mazzia\inst{2} \and Marc Ciligot-Travain\inst{4}}
\authorrunning{S. Madmar et al.} 
%


\institute{ESPACE, CNRS, Avignon University,  Avignon, France\\
\email{sofiane.madmar@cnrs.fr}
\and
IMBE, Avignon Univ, Aix Marseille Univ, CNRS, IRD, Marseille, France\\
\and 
LIA,  Avignon University,  Avignon, France
\and
LMA, Avignon University,  Avignon, France}

\maketitle              

\begin{abstract}
Researchers worldwide have drawn inspiration from nature to optimize network design and dynamics. Some of the wonders of the living world exhibit remarkable abilities in generating efficient and resilient spatial structures. By mimicking biological strategies, transportation infrastructures could be profoundly rethought. This paper aims to provide the basis for a biomimicry framework for addressing transportation networks. In light of examples from the literature, the relevance of such a framework for advancing research in nature-inspired networks is demonstrated, with the aim of achieving resilience and efficiency.
\keywords{biomimicry, transportation network, optimization, robustness, resilience}
\end{abstract}
%

\section{Introduction}
\label{sec:intro}
 A transportation network is a system of connected points, such as road intersections or bus stops, and the paths between them, such as streets or rail lines, that allow for the movement of people and vehicles across an area~\cite{bell_transportation_1997}. When inefficient, it can lead to congestion, supply chain failures, and abrupt modal shifts---all of which trigger cascaded failures with significant socio-economic consequences. The expansion of networks driven by urban sprawl represents an other fundamental challenge for network science, as it necessitates a delicate trade-off between maximizing accessibility, overall efficiency, and cost-effectiveness of the system. Ultimately, transportation networks are a primary determinant of territorial resilience, which means the capacity to anticipate, respond to, and adapt to systemic shocks, ensuring sustainable development regardless of perturbations~\footnote{ https://www.cerema.fr/system/files/documents/2020/10/boussoleresilience-cerema-web-finalpdf.pdf}. For all these reasons, transportation networks exert a profound influence across all sectors of society. 

Addressing these socio-economic and territorial vulnerabilities requires a paradigm shift toward more sustainable and resilient design strategies, that is ongoing~\cite{filippi_paradigm_2022}. \textit{Biomimicry} offers a solid framework for this transition by drawing inspiration from the living world to find innovative solutions for human-centred challenges. It seeks to emulate functions and strategies from nature, namely, it may imply an abstraction of the biological subject~\cite{benyus_biomimicry_1997}. According to the related ISO standard\footnote{\url{https://www.iso.org/obp/ui/fr/#iso:std:iso:18458:ed-1:v1}}, biomimicry encompasses ``philosophy and interdisciplinary design approaches taking nature as a model to meet the challenges of sustainable development (social, environmental, and economic)''. 

According to Benyus (1997), the foundations of biomimicry can be summarized in three main principles~\cite{benyus_biomimicry_1997}: 
(i) ``Nature as Model'', which establishes the imitation of nature's patterns and strategies to respond to technological challenges; 
(ii) ``Nature as Measure'', which establishes nature as a standard to validate or reject a model according to its environmental impacts and ecological fit; and 
(iii) ``Nature as Mentor'', described by Dicks (2018) as an epistemological principle which defines nature as a direct source of knowledge through observation~\cite{dicks_nature_2018}. This approach not only fosters sustainable design~\cite{raman_mapping_2024} but also encourages transdisciplinarity, which is essential for addressing the complex challenges of our era~\cite{madmar_applications_2023}. We base our paper on this epistemological framework 
and call it \textit{Benyusian biomimicry} as previously done~\cite{madmar_applications_2023}, in order to distinguish it from any other ways to understand biomimicry. 

Network science is not unfamiliar with inspiration from nature, particularly in the last 20 years, mainly through the application of natural computing and multi-agent systems~\cite{gerbaud_bibliometric_2022}, for tasks such as network design, travelling salesman problem, scheduling and routing~\cite{gerbaud_bibliometric_2022,kar_bio_2016}. However, these works rely on the notion of \textit{bioinspiration}, which is different from biomimicry. According to the above-mentioned ISO standard, it is defined as a ``creative approach based on the observation of biological systems''. 
According to Scopus statistics\footnote{We used the following query: \textit{TITLE-ABS-KEY ( "biomim*" OR "bio-inspir*" OR "bioinspir*" OR "nature-inspired" ) AND TITLE-ABS-KEY ( "network science" OR "network theory" OR "graph theory" OR "complex networks" )}}, among 1,115 network science-relate documents dealing with inspiration from nature, only
9 of them refer directly to ``biomimicry'' in their keywords. This shows that network science research has made very little use of \textit{Benyusian biomimicry} concepts. 

The distinction between biomimicry and bio-inspiration is crucial. One the one hand, \textit{Benyusian biomimicry} considers nature as a direct source of knowledge and a normative standard~\cite{blok_ecological_2016}. It has developed a comprehensive methodology that directs inspiration drawn from nature toward sustainable development. One the other hand, if bioinspiration includes the core principle of ``Nature as Model'', it does not rely on a specific method. Moreover, without considering ``Nature as Measure'', bioinspired solutions---though they may extrinsically achieve sustainable results---do not inherently address sustainability~\cite{benyus_biomimicry_1997,blok_ecological_2016,dicks_nature_2018}. Finally, contrary to the ``Nature as Mentor'' principle, if nature is perceived as a mere ``starting point'' for innovation rather than as a mentor, then the iterative development process tends to distance itself from biological reality, which then becomes, at best, a shortcut to innovation, and at worst, a pretext~\cite{dicks_nature_2018}. 
Thus, \textbf{by overlooking biomimicry, research in transport networks may dismiss a critical framework that prioritizes long-term sustainability over immediate efficiency and integrates iteratively biological strategies for resilient transport networks.}

In this paper, we aim at showing concretely that  \textit{Benyusian biomimicry} constitutes a more comprehensive conceptual framework than bioinspiration to deal with transport network research. 
We consider these networks at two levels: on the one hand, the methods behind the generation of the network, and its resulting structural properties, and on the other hand, the processes that occur on the network, in particular those related to path optimization. 
Our contribution is threefold. First, we discuss \textbf{how the ``Nature as Model'' concepts can be translated into actionable tools for transport network tasks}, and provide examples of this process. Second, \textbf{we leverage ``Nature as Measure'' to design transport network-specific means of assessing the resulting models}. Third, we discuss \textbf{the relevance of ``Nature as Mentor'' as a way to avoid overlooking valuable biological functions. }

The rest of the article is organized as follows. Section~\ref{sec:NatAsModel} focuses on models from nature. Section~\ref{sec:AbsModelling} shows various ways to abstract knowledge from nature. Lastly, Section~\ref{sec:Discussion} is a discussion about nature as a measure and as a mentor for network design. 
We conclude in Section~\ref{sec:conclu}, proposing a visual summary of the framework.

\section{Nature as a Model for Transportation Networks}
\label{sec:NatAsModel}
This section details the application of the principle of ``Nature as Model''. We initially introduce several significant biological models proposed in the literature to enhance network design or address optimization issues. Subsequently, we outline guidelines for determining an appropriate model

\subsection{Biological Models}
\label{subsec21}
Biomimicry has three levels of application---form, process, and ecosystem---covering a wide range of problems. We explore here some biological strategies that have been suggested as models for transportation networks in the literature. To our knowledge, there are no existing works at an ecosystem level, which would mean to learn from the systemic functioning in nature. Consequently, only the form and process levels—--and their combinations—--are covered. This level sorting provides a foundation for evaluating the appropriateness of a model.

\subsubsection{Form Level.}
\label{subsec22}
In nature, certain morphological structures inspired transportation network design. The structures of living organisms which exhibit functional topologies can be a source of inspiration for better understanding what governs the robustness~\footnote{The ability of a network to sustain its normal functionality when a fraction of the network fail to work due to attacks~\cite{lou_structural_2023}.} of a network. Certain forms are particularly effective at fulfilling specific functions, such as the distribution of fluids in plant tissues through their veins, which simultaneously provide mechanical resistance~\cite{hong_trade-off_2025}. Drawing inspiration from forms does not mean simply copying the shape of a leaf, but rather understanding how its topology makes it functional and long-lasting, consequently, drawing lessons for transportation networks.

In this regard, orb webs are an interesting example. These gems of natural architecture are produced by certain spiders as both a habitat and a predation tool. The orb web geometric structure results in a good dissipation of a prey collision kinetic energy even if some threads are broken. When a pray is captured in the sticky silk, its vibrations inform the spider with utmost efficiency due to longitudinal transmission along the radii toward the centre. The mesh properties have also an effect on its ability to cover a maximum of area in the capture area while optimizing material cost~\cite{sensenig_mechanical_2013}. The spider maintains a less dense zone---comprising only radial threads---between the central hub and the peripheral capture area. This ``free zone'' functions as both a structural and functional buffer: it isolates vibrations to enhance prey detection and provides a non-sticky corridor for rapid traversal.~\cite{sensenig_mechanical_2013,wirth_forces_1992}.

\subsubsection{Process Level.}
While the form level focuses on the general topological properties of a network, the process level mainly addresses contextual optimization. Indeed, various organisms and biological processes have served as sources of inspiration for solving optimization problems and developing transportation systems. Researchers have identified behaviours and dynamics in natural group of organisms that have been proven highly efficient when abstracted into metaheuristic algorithms and multi-agent systems~\cite{blum_ant_2005,del_ser_bioinspired_2020}.

At this level, the ant colony is a prominent model. Ants are eusocial\footnote{Social organization of animals whereby the community is divided into castes of fertile and infertile individuals.} insects that work collectively to thrive and grow. Ant colonies have no centralized command. Instead, they rely on indirect communication---known as \textit{stigmergy}---using pheromone trails to share information. When foraging, ants secrete pheromones to provide information about the proximity and quality of a food source~\cite{schmid-hempel_b_2002}. Through the accumulation of pheromones and the evaporation of trails on longer, less efficient paths, the colony optimizes its travel time to the food~\cite{deneubourg_self-organizing_1990}. There is a double level of observation: working insects foraging and leaving pheromones, and the global colony that operates as a whole, exploiting optimal paths built on deposits.

\subsubsection{Mixed Levels.}
An other well-studied organism to mention is \textit{Physarum Polycephalum}, an acellular \textit{amoebozoa} slime mould that lives in dead leaves and tree bark from wet forests. In its plasmodium phase, it uses a cytoplasmic streaming to move and forage for food. By gradually refining its tubular network upon reaching food, it creates cytoplasmic networks with stupendous properties. Nakagaki \textit{et al.} (2000) have shown that \textit{P. Polycephalum} is showing a primitive intelligence in resolving a maze and finding the shortest path~\cite{nakagaki_maze-solving_2000}. Further works have analysed the network structure of \textit{P. polycephalum} in the case of multiple food sources and describe the plasmodium's network as a combination of Steiner's minimum tree and cyclic connections; this induces a balance between optimality (minimal cytoplasm cost) and robustness against random or targeted disruptions~\cite{nakagaki_obtaining_2004}.

The biological model here is both at the form and process levels: the way in which the plasmodium expands and fine-tunes its network is related to processes, whereas its tubular networks global topological properties relates to forms.

Under the hypothesis that survival is the primary biological imperative, process-level models may inherently operate according to a logic of resilience. In such cases, the selection of a path is not merely a single-objective search for performance, but also a strategic choice implied by the need of a robust topology that ensures long-term systemic stability.

\subsection{How to Evaluate a Model's Relevance?}
\label{subsec22}

Once a biological model has been identified, its appropriateness must be rigorously assessed in relation to the specific problem at hand. We propose two criteria to evaluate the relevance of these models, building upon \textit{Benyusian biomimicry}. Through the previous biological examples, we demonstrate how this approach clarifies their practical application to transportation networks.

\subsubsection{\textit{Criterion 1: Functional Analogy.}}
\label{subsubsec221} 
\begin{figure}[h]%
\centering
\includegraphics[width=0.7\textwidth]{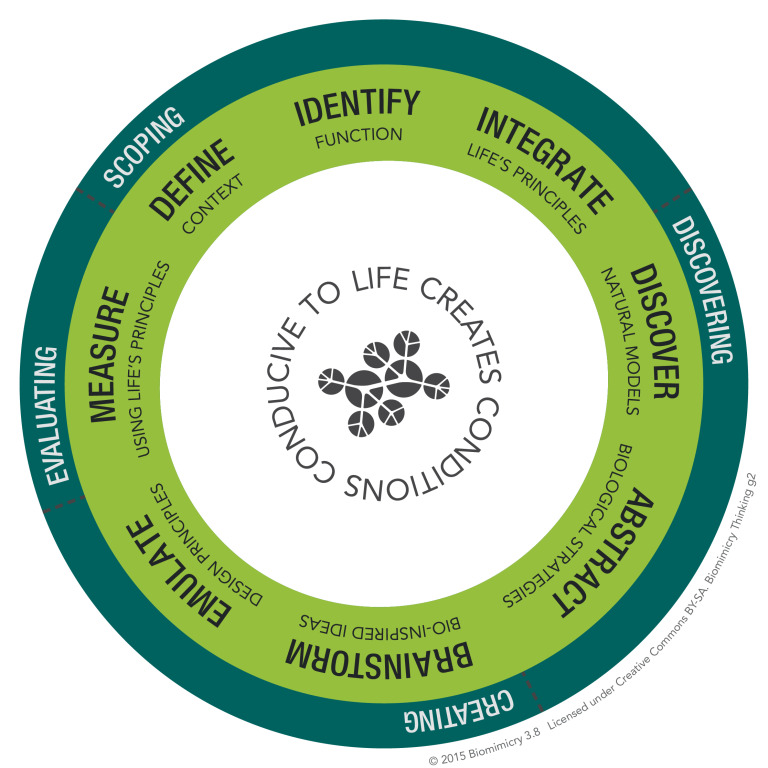}
\caption{Biomimicry DesignLens (Biomimicry 3.8)}\label{fig1}
\end{figure}

In Benyus's co-funded institutions such as \textit{Biomimicry 3.8} with their \textit{DesignLens}\footnote{\url{https://biomimicry.net/the-buzz/resources/biomimicry-designlens/}} which is a collection of diagrams that summarize the design methodologies and principles of \textit{Benyusian biomimicry} (cf. Figure~\ref{fig1})
and \textit{Ask Nature} website\footnote{\url{https://asknature.org/}}, the main input for selecting an appropriate model is its strategy. Such strategies are organized by groups and subgroups in a ``biomimicry taxonomy'' resulting in a list of specific functions. This requires a rigorous functional analogy between the biological model and the intended technical application. Within transportation networks, these functions are diverse and can be categorized by scale. At the optimization level, objectives include distance minimization, throughput maximization, congestion control, travel time reduction. Conversely, at a systemic level, the focus shifts to global network properties, such as robustness to disruptions, territorial coverage, accessibility, infrastructure cost or topological straightness~\footnote{Ratio between the Euclidean
distance and the actual network distance.}.

However, these functions do not operate in isolation. It is frequently needed to solve multi-objective problems and/or bi-level interactions. In the case of \textit{P. polycephalum}, we could describe the network generation through an iterative feedback loop: the cytoplasmic flows optimize for local nutrient transport, which in turn reinforces the tubular morphology into a robust structure~\cite{adamatzky_bioevaluation_2012,hausler_conception_2012,nakagaki_obtaining_2004,tero_physarum_2006}.

\subsubsection{\textit{Criterion 2: Contextual Fit.}}
\label{subsubsec222}
Even if a given problem can be solved by a biological function, it does not mean it is relevant to use it in a specific context. For instance, the orb web morphology has multiple functions leading to properties: straightness, cost-optimum spatial coverage and accessibility. However, it is not clear in which context to use these properties. Is it relevant to use this model to design a city's road network? Can we use it for public transportation networks? We have to question the context of application of a natural strategy. It is why the Biomimicry 3.8 DesignLens teaches to define the context and accordingly identify the right function. 
For instance, we suggest that the free zone of the orb web can contribute to discussions on the context of transitions between urban centres and suburbs because of its function of channelling information (that can be modelled to energy or traffic) and freeing the in-between space.
Therefore, it is necessary to consider the applicability of a biological strategy to a problem within networks. This may even require a significant level of abstraction, given important constraints such as the scale of application and the cost of infrastructure.

\section{Abstracting and Modelling Life: Complexity Science}
\label{sec:AbsModelling}

The dialogue between life sciences and transportation engineering appears to be effective through the tools of complexity science. We summarize in this section various ways of translating biological knowledge into technological models for transport networks. We refer to the biological models mentioned in Section~\ref{sec:intro}.

\subsection{Graph Theory}
\label{subsec31} 
Using a graph to model a biological network and analyse its proprieties is the primary approach for mimicking the biological functions of a form level model. For instance, Josselin \textit{et al.} (2015) have studied the properties of orb webs by modelling them through a graph network and compared it to lattice and Manhattan (grid) networks~\cite{josselin_modelisation_2015}. It seems that the orb web, due to its radial and spiral threads, shows better straightness. Thus, spatial coverage and accessibility are enhanced with optimized material use, making up a very efficient hub~\cite{josselin_modelisation_2015,josselin2016}. Graphs serve as highly abstracted intermediaries that facilitate the transdisciplinary translation between biological phenomena and technical applications. By reducing complex organic structures to their topological essentials (nodes and edges), graph theory provides a common mathematical framework to evaluate biological phenomena within technical applications.

\subsection{Metaheuristic Algorithms}
\label{subsec32}
For process level biomimicry, the use of metaheuristic problem-solving algorithms provides an important bridge between behavioural observations and computer simulations. A metaheuristic is a high-level strategy that guides search algorithms to find near-optimal solutions for complex problems by balancing broad exploration with targeted refinement. For instance, Dorigo (1992) theorized a metaheuristic inspired by ants' pheromone-based foraging behaviour in his thesis~\cite{dorigo_optimization_1992}. This marked the birth of the Ant Colony Optimization (ACO) research domain, which has seen further development~\cite{blum_ant_2024}. Even if it can be considered oversimplified in front of real ants' complex behaviour, ACO has proven its efficiency as a metaheuristic for resolving optimization problems such as Travelling Salesman Problem (TSP), Steiner Tree Problem (STP), and many others including routing problems. This model has been enhanced and modified for very specific purposes. ACO is mainly used for solving combinatorial optimization problems and it is still gaining interest in research papers~\cite{blum_ant_2024}. 

\subsection{Other Tools: Physics and Multi-Agent Simulation}
\label{subsec33}
Regarding \textit{P. Polycephalum}, the literature show a variety of modelling to mimic its foraging behaviour. The first model was based on fluid dynamics (Poiseuille flow)~\cite{tero_rules_2010} and others followed such as cellular automaton, reaction-diffusion of Belousov-Zhabotinsky, swarm particle multi agent model, and more~\cite{awad_survey_2022}. Overall, these are tools of physics that can be used to abstract the knowledge from nature.

Additionally, an other approach is multi-agent system. By integrating behavioural discoveries into interaction between agents, such a system can model more complex dynamics to travelling along a network and at the same time better understand how its structure influences individual and collective choices. For instance, it has been used for the dynamic optimization of transport services~\cite{lammoglia_analyse_2013}, and for various problems in transportation networks~\cite{del_ser_bioinspired_2020}.

\section{Measure and mentor: a Discussion}
\label{sec:Discussion}

As established in the Section~\ref{sec:intro}, the normative and epistemological principles constitute the core essence of biomimicry. In this section, we discuss the concept of biorationality as a potential specific methodology for ``Nature as Measure'' applied to networks. Then we show why it is important to hold on the ``Nature as Mentor'' pillar in the light of the ant colony model.

\subsection{Toward a Biorationality?}
\label{subsec41}

Bridging the gap between the need for resilient transportation networks and engineering solutions requires a rigorous evaluation framework. Within the biomimicry paradigm, the dialectical interplay between modelling and measurement ensures the reliability of the knowledge creation process~\cite{dicks_nature_2018}. The superior model is distinguished by its capacity to align with the biological measure, thereby ensuring that the design strictly adheres to natural normative standards.
``Nature as Measure'' usually involves assessing designs against ``Life's Principles''---general patterns observed across ecosystems that makes them sustainable---in order to maintain the ``conditions conducive to life'' as described in the DesignLens\footnote{\url{https://biomimicry.net/the-buzz/resources/designlens-lifes-principles/}}. Note that it can also be used as a normative tool to evaluate models that are not biological. 

Despite already indirectly covering main transportation challenges such as pollution, land use, and congestion, to our knowledge there does not seem to be any formalized nature-based network evaluation methodology that addresses topology in the literature. However, recent research on \textit{P. polycephalum} is promising.

Specifically, Tero \textit{et al. }(2010) showed that the Tokyo railway network exhibits efficiency (minimum distance), fault-tolerance to accidental disconnections, and total material costs comparable to the biological networks formed by the slime mould when connecting an analogous set of food sources in only a few hours~\cite{tero_rules_2010}. Similarly, Adamatzky \textit{et al. }(2012) compared international road networks with slime mould patterns, using metrics such as average path length, diameter, average degree, number of independent cycles and the Harary, $\Pi$, and Randić indices~\cite{adamatzky_bioevaluation_2012}.
While their concept of ``biorationality''---the degree of convergence with biological results---provides a promising foundation as a measuring methodology, we suggest that a comprehensive methodology requires a combination of these geometric and topological metrics with formal evaluations of robustness and cost.

\subsection{How not to Overlook Nature's Knowledge?}
\label{subsec42}

To complete biomimicry pillars, we transition from model and measure to the contribution of ``Nature as Mentor'', exemplified by ant colonies. Research in ACO have mostly leaned toward mathematical abstraction, often detaching itself from its biological roots. Just as research in aviation is looking back to birds for improvements~\cite{fluck_lifting_2014}, we may consider focusing again at the ant model to instigate new design paradigms. 
Indeed, recent ethological studies reveal sophisticated congestion management capabilities previously overlooked in metaheuristic research~\cite{guerrieri_anti-jam_2025,poissonnier_experimental_2019}. Notably, the work of Czaczkes \textit{et al. }(2011) on \textit{Lasius niger} showed a synergy between collective pheromone trails and private navigational memory. This dual-information system significantly reduces ``U-turns''---a proxy for navigational uncertainty---thereby increasing foraging efficiency. Overall, these results are suggesting various behaviours can improve ants' collective mobility, that are not covered by ACO and may be modelled through multi-agent simulations rather than classic metaheuristic algorithms. 
Crucially, a gap remains in the inspiration (form level) of complex networks made by ants between multiple sources or multiple nests, although its efficiency has been shown to be comparable to human networks~\cite{buhl_shape_2009} and presenting robustness in its topology~\cite{cook_efficiency_2014}. Namely, ant colony could also be understood as a mixed level model.

To remain faithful to the principle of ``Nature as Mentor'', biomimicry must be understood as an iterative process that does not oversimplify biological properties in the pursuit of improving transportation networks. Otherwise, we risk overlooking vital insights derived from biological necessities, such as foraging efficiency and adaptation to environmental uncertainties.

\section{Conclusion}
\label{sec:conclu}

Despite its critical role, the network as a distinct object of study remains significantly under-explored in research on transportation. This lack of scrutiny raises questions about the structural effects of network structures on transport practices and whether certain configurations inadvertently encourage urban sprawl~\cite{foltete_impacts_2008}. 

\begin{figure}[h]%
\centering
\includegraphics[width=1\textwidth]{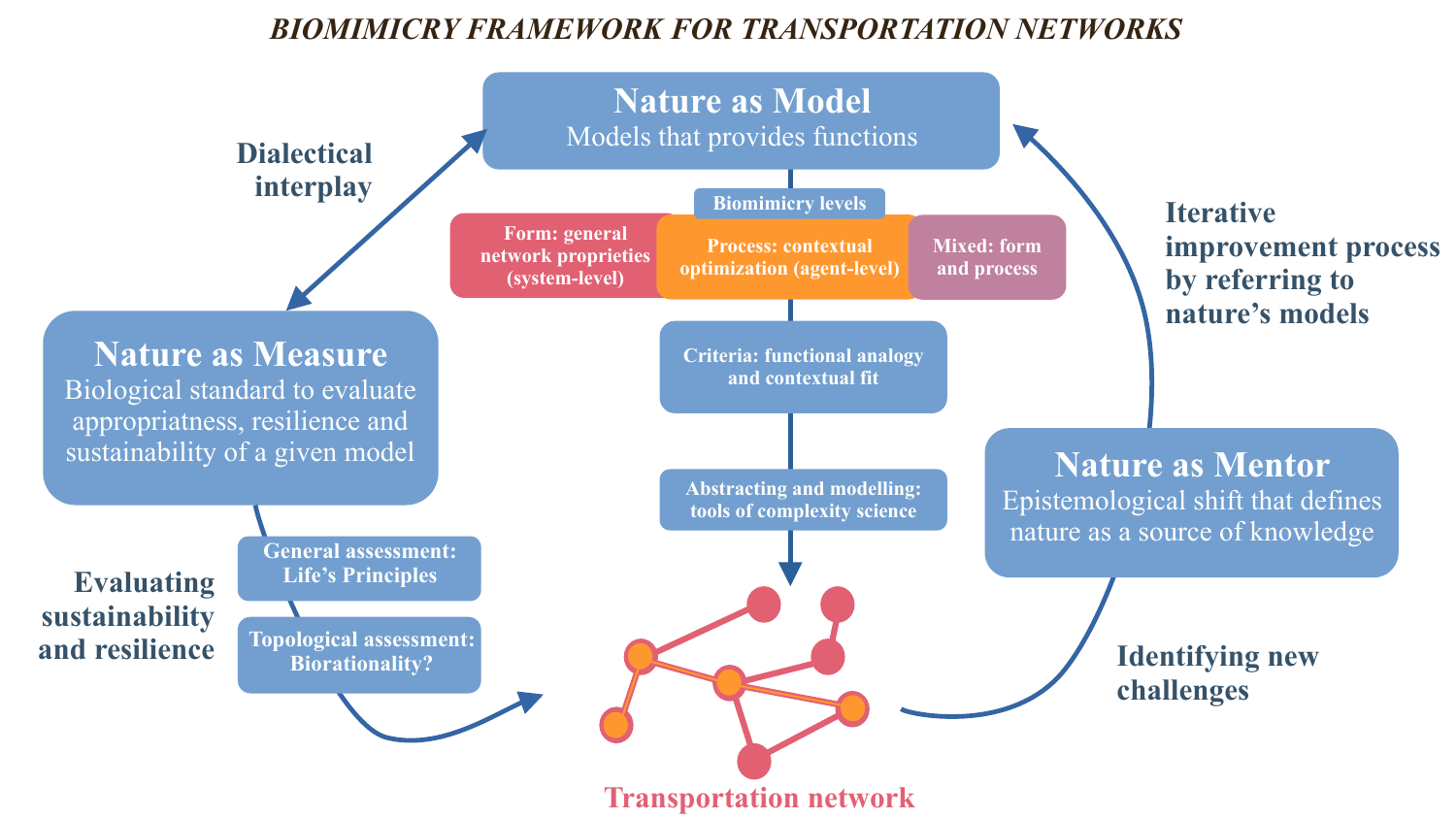}
\caption{Biomimicry framework for transportation networks}\label{fig2}
\end{figure}

In nature, organisms are always facing multiple constraints which result in topological trade-off between foraging efficiency and systemic robustness to environmental uncertainties. Nature thus serves as a profound source of inspiration for deciphering the complex dynamics inherent in human networks. In this sense, biomimicry represents a methodology that allows for identifying appropriate models that can be abstracted through various approaches. Our biomimicry framework is summarized in Figure~\ref{fig2}. The central vertical axis illustrates the modelling steps derived from nature, as described in Section~\ref{sec:NatAsModel}.

Biological organisms and processes offer insights that transcend mere design enhancement. These models serve as a critical lens, challenging the efficiency-centric focus of existing infrastructures and exposing their inherent vulnerabilities. By re-defining the network as a vital component of sustainability, natural patterns question our singular emphasis on path minimization and enhance our capacity to cultivate systemic resilience through network robustness. This principle imposes a systemic vision as opposed to a reductionist vision purely focused on local performance. The left side of our schematic framework depicts these aspects through the model evaluation approaches described in Section~\ref{sec:Discussion}.

Finally, by fully grasping the potential of this approach, we recognize that nature is a direct source of epistemic value. This shift transforms biological research into a powerful catalyst for innovation, providing the foundations upon which a resilient future for transportation can be built. When identifying new challenges, researchers can refer back to the biological "library of solutions" provided by nature. This idea is displayed in the right side of the schema showing that ``Nature as mentor'' is essential in looking back to biological models.

Ultimately, integrating \textit{Benyusian biomimicry} framework in transportation networks allows to better acknowledge the benefits of inspiring from nature while relying on a clear methodology.
%

\end{document}